\documentclass[a4paper,11pt]{article}
\pdfoutput=1 

\usepackage{jinstpub} 

\usepackage{pdflscape}
\usepackage{afterpage}
\usepackage{adjustbox,lipsum}

\usepackage{graphicx}
\usepackage{dcolumn}
\usepackage{bm}
\usepackage{amsmath}
\usepackage{diagbox}
\usepackage{rotating}
\usepackage{multirow}
\usepackage{xcolor}
\usepackage{hepunits}
\usepackage{multirow}

\usepackage{hyperref}
\usepackage[mathlines]{lineno}

\title{Measurement of the bulk radioactive contamination of detector-grade silicon with DAMIC at SNOLAB}

\author[a]{A.~Aguilar-Arevalo} 
\affiliation[a]{Universidad Nacional Aut{\'o}noma de M{\'e}xico, Mexico City, Mexico} 

\author[b]{D.~Amidei}
\affiliation[b]{Department of Physics, University of Michigan, Ann Arbor, Michigan, United States}  

\author[c]{D.~Baxter}
\affiliation[c]{Kavli Institute for Cosmological Physics and The Enrico Fermi Institute, The University of Chicago, Chicago, Illinois, United States}

\author[d]{G.~Cancelo}
\affiliation[d]{Fermi National Accelerator Laboratory, Batavia, Illinois, United States}

\author[a]{B.A.~Cervantes Vergara}

\author[e]{A.E.~Chavarria}
\affiliation[e]{Center for Experimental Nuclear Physics and Astrophysics, University of Washington, Seattle, Washington, United States}

\author[c]{E.~Darragh-Ford}

\author[a]{J.C.~D'Olivo}

\author[d]{J.~Estrada}

\author[a]{F.~Favela-Perez}

\author[f]{R.~Ga\"ior}
\affiliation[f]{Laboratoire de Physique Nucl\'eaire et des Hautes \'Energies (LPNHE), Sorbonne Universit\'e, Universit\'e de Paris, CNRS-IN2P3, Paris, France}

\author[d,1]{Y.~Guardincerri\note{Deceased January 2017}}

\author[g]{T.W.~Hossbach}
\affiliation[g]{Pacific Northwest National Laboratory (PNNL), Richland, Washington, United States} 

\author[h]{B.~Kilminster}
\affiliation[h]{Universit{\"a}t Z{\"u}rich Physik Institut, Zurich, Switzerland }

\author[i]{I.~Lawson}
\affiliation[i]{SNOLAB, Lively, Ontario, Canada }

\author[h]{S.J.~Lee}

\author[f]{A.~Letessier-Selvon}

\author[c,f]{A.~Matalon}

\author[e]{P.~Mitra}

\author[e]{A.~Piers}

\author[c,f]{P.~Privitera}

\author[c]{K.~Ramanathan}

\author[f]{J.~Da~Rocha}

\author[a]{Y.~Sarkis}

\author[j]{M.~Settimo}
\affiliation[j]{SUBATECH, CNRS-IN2P3, IMT Atlantique, Universit\'e de Nantes, Nantes, France}

\author[c]{R.~Smida}

\author[c]{R.~Thomas}

\author[d]{J.~Tiffenberg}

\author[f]{M.~Traina}

\author[k]{R.~Vilar}
\affiliation[k]{Instituto de F\'isica de Cantabria (IFCA), CSIC--Universidad de Cantabria, Santander, Spain}

\author[k]{A.L.~Virto}

\emailAdd{amatalon@uchicago.edu}

\abstract{We present measurements of bulk radiocontaminants in the high-resistivity silicon CCDs from the DAMIC experiment at SNOLAB. We utilize the exquisite spatial resolution of CCDs to discriminate between $\alpha$ and $\beta$ decays, and to search with high efficiency for the spatially-correlated decays of various radioisotope sequences. Using spatially-correlated $\beta$ decays, we measure a bulk radioactive contamination of $^{32}$Si in the CCDs of $140 \pm 30$ $\mu$Bq/kg, and place an upper limit on bulk $^{210}$Pb of $< 160~\mu$Bq/kg. Using similar analyses of spatially-correlated $\alpha$ and $\beta$ decays, we set upper limits of $< 11$ $\mu$Bq/kg (0.9 ppt) on $^{238}$U and $< 7.3$ $\mu$Bq/kg (1.8 ppt) on $^{232}$Th in the bulk silicon. The ability of DAMIC CCDs to identify and reject spatially-coincident backgrounds, particularly from $^{32}$Si, has significant implications for the next generation of silicon-based dark matter experiments, where $\beta$'s from $^{32}$Si decay will likely be a dominant background. 
}

\keywords{Dark Matter detectors (WIMPs, axions, etc.); Solid state detectors; Search for radioactive and fissile materials; Particle identification methods}

\arxivnumber{2011.12922} 

\collaboration{DAMIC Collaboration}

\begin{document}
\maketitle
\flushbottom

\section{Introduction}
There is overwhelming astrophysical and cosmological evidence for Dark Matter (DM) as a major constituent of the universe~\cite{DM-history}, in particular by how its gravitational presence affects dynamics of galaxy clusters~\cite{Zwicky}, galactic rotation curves~\cite{Rubin}, and features of the Cosmic Microwave Background~\cite{Planck}. DM comprises about 27\% of the energy density of the universe; ordinary matter comprises 5\%. Determining the so-far-elusive nature of DM remains one of the most important scientific efforts today. One way to probe the nature of DM is to measure whether it couples to standard model particles through a non-gravitational interaction. Measuring the energy deposition from such an interaction in a particle detector is what we refer to as direct-detection~\cite{goodman,jungman,tasi}.

One of the largest challenges in constructing a direct-detection experiment is that DM is not the only source of energy depositions in the detector. Other sources, predominantly radiogenic backgrounds, can mimic a DM signal and fundamentally limit the sensitivity of a direct-detection experiment to DM models. The primary means of addressing such backgrounds are mitigation and rejection. Mitigation includes material selection and shielding. Detector materials are carefully selected and measured to ensure low levels of long-lived radioactive isotopes, such as primordial $^{238}$U and $^{232}$Th. Rejection is the process by which radioactive backgrounds are eliminated using analysis techniques to distinguish standard model decays from a potential DM signal. With this analysis, we show how the capabilities of silicon charge coupled devices (CCDs) can identify radioactive background events to enable and guide both techniques.

We present improved measurements of bulk radioactive backgrounds in  high-resistivity silicon CCDs from the DArk Matter In CCDs (DAMIC) experiment at SNOLAB. We build on a method to distinguish and reject background events coming from the same radioactive decay chain, initially presented in Ref.~\cite{Contamination}. By utilizing the precise spatial resolution of CCDs and discrimination between $\alpha$ and $\beta$ particles, we identify spatially-correlated radioactive decay sequences over periods of up to several weeks. In addition to constraining $^{238}$U and $^{232}$Th activities with comparable sensitivity to mass-spectroscopy techniques~\cite{icpms}, we employ this method to measure, with unprecedented sensitivity, shorter-lived isotopes in the bulk silicon, such as $^{32}$Si and $^{210}$Pb.

\begin{figure}[t]
\begin{center}
\includegraphics[width=0.7\textwidth, trim=10 0 0 60, clip=true]{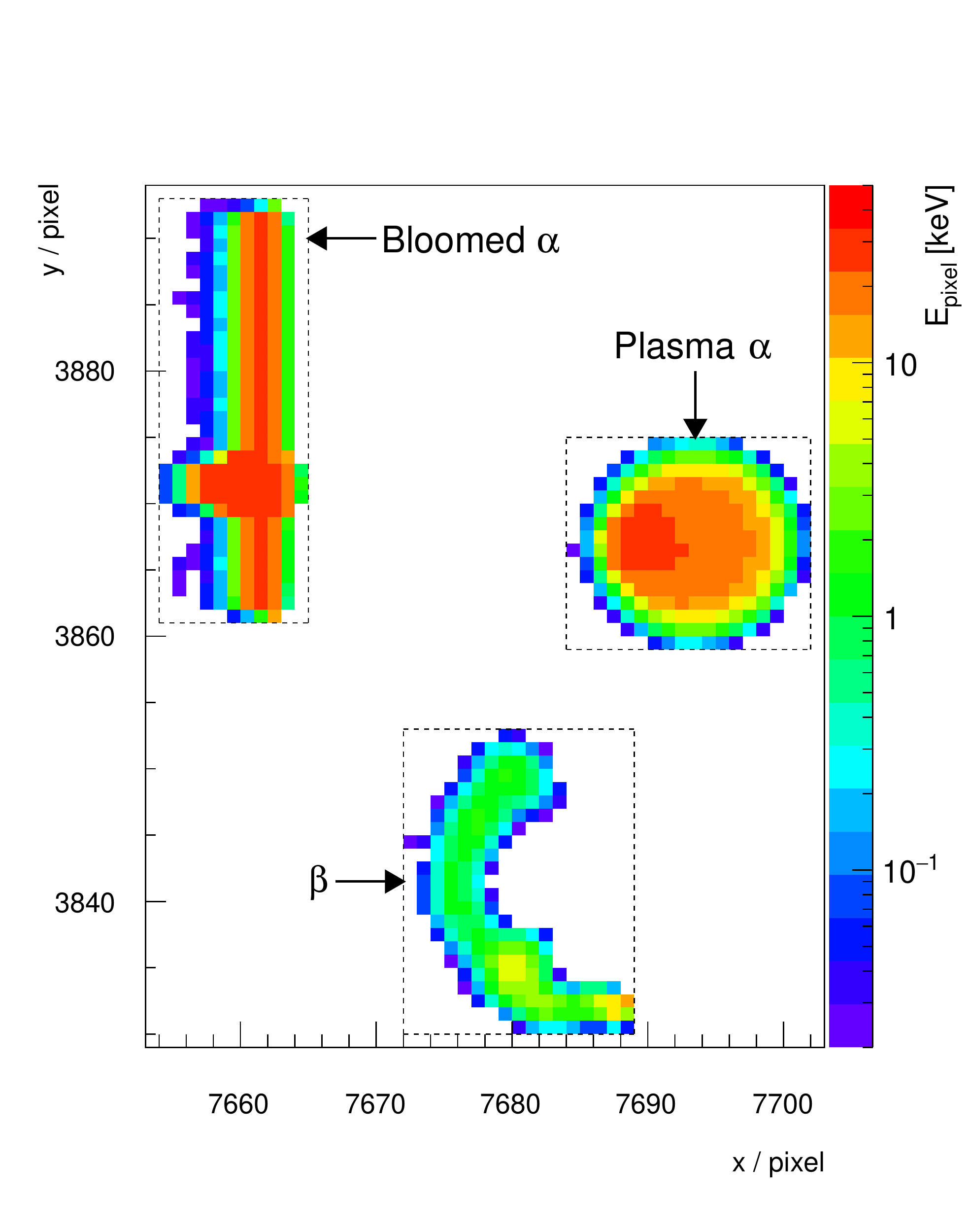}
\caption{Example of $\alpha$ and $\beta$ clusters reconstructed in data. We note the bloomed $\alpha$'s higher spatial variance in  $\hat{y}$ compared to that of the plasma $\alpha$ (see text). The ``worm"-like track of the $\beta$ takes up little space inside the rectangle drawn around it.}
\end{center}
\label{fig:events}
\end{figure}

By virtue of the low readout noise of CCDs and relatively low mass of the silicon nucleus, DAMIC is especially sensitive to low mass ($<10$~GeV/$c^{2}$) weakly interacting DM~\cite{damic2020}. The experiment is located 2 km underground in the SNOLAB laboratory and uses a tower of 16-Mpix CCDs, 
which have been acquiring data since 2017. A detailed discussion of the DAMIC detector apparatus can be found in Ref.~\cite{damic2020}. The DAMIC CCDs were developed at LBNL MicroSystems Lab and feature a 4116$\times$4128 array of pixels 15$\mu$m$\times$15$\mu$m in size with a total thickness of $674 \pm 3$\,$\mu$m~\cite{LBNL}. 
Each CCD is constructed from Topsil high-resistivity ($>10$~k$\Omega$~cm) float-zone silicon and is fully depleted by the application of 70\,V to a thin backside planar contact, which leads to an active thickness of $665 \pm 5$\,$\mu$m. 

Charge from ionization is drifted along the direction of the electric field ($-\hat{z}$) and experiences lateral thermal diffusion as it transits the bulk material. The charge carriers (holes) are collected and held $<1~\mu$m below the gates for the duration of a long exposure until readout. The digitized saturation value for this data is higher than the pixel full-well capacity in order to ensure all charge is measured for high-energy $\alpha$'s. During readout, charge is transferred laterally from pixel to pixel along the $x-y$ plane ($z=0$) by modulating the gate electrodes, which include ``parallel clocks" in $\hat{y}$ and higher frequency ``serial clocks" in $\hat{x}$. Parallel clocks transfer charge into the serial register, and serial clocks transfer charge from the serial register to the CCD's output node, where it is measured by a correlated double-sampling circuit, described in Ref.~\cite{Janesick}. 
The results of this technique are high-resolution images with low pixel readout noise $\sigma_{\textrm{pix}}$ (found to be as low as 2e$^-$~\cite{damic2020}) containing a two-dimensional projection of all ionization events that occurred during the exposure. For this analysis, we set a conservative analysis threshold of 500 eV. 

\section{Methodology}
Background data used in this analysis was acquired between February and September 2017 with six CCDs (total exposure time $t$ =  181.3~d; total mass $M$ = 36~g). A seventh CCD began taking data partway through the time period of this analysis, and is thus excluded. The energy response of each CCD was obtained during the commissioning phase at SNOLAB, similar to the procedure in Ref.~\cite{0.6}. Images in this dataset had $t_{\textrm{exp}} = 3 \times 10^{4}$ s (0.35 d) exposure and pixels were read out individually to improve spatial resolution. Images are processed using the same procedure described in Ref.~\cite{0.6}, including steps for pedestal subtraction, correlated noise subtraction, and masking regions with high leakage current or defects. After image processing is completed, clustering reconstruction is performed: contiguous pixels not excluded by masked regions are grouped together if they exceed a threshold value of 4$\sigma_{\textrm{pix}}$. 

\begin{figure}[t]
\begin{center}
\includegraphics[width=0.7\textwidth, trim=0 0 30 0, clip=true]{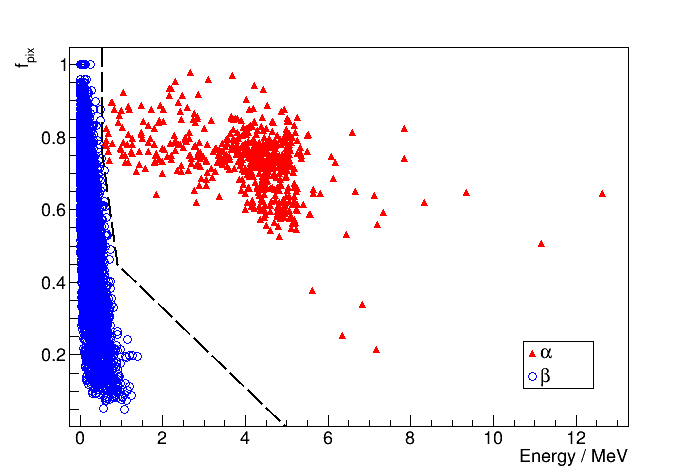}
\includegraphics[width=0.7\textwidth, trim=0 0 30 10, clip=true]{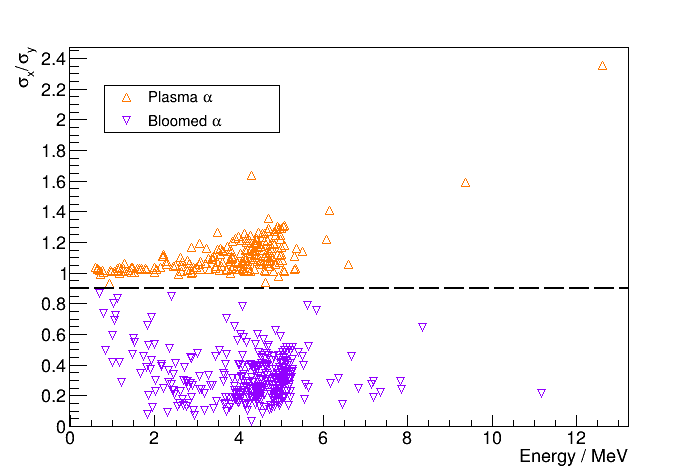} 
\end{center}
\caption{Top: discrimination of $\alpha$'s (triangles) and $\beta$'s (circles) in the 2017 background dataset using $f_{\textrm{pix}}$, the fraction of a cluster's pixel number over the smallest rectangle covering its area. Bottom: discrimination of bulk or back surface (plasma) $\alpha$'s from front surface (bloomed) $\alpha$'s using the ratio $\sigma_{x}/\sigma_{y}$, which leverages bloomed $\alpha$'s higher spatial variance in $\hat{y}$.}
\label{fig:discrimination}
\end{figure}

In DAMIC CCDs, low-energy electrons and nuclear recoils produce clusters whose spatial extent is due primarily to charge diffusion. The spatial characteristics of reconstructed $\alpha$ and high-energy $\beta$ clusters enable efficient event-by-event discrimination. In particular, $\beta$'s have long, ``worm"-like tracks compared to $\alpha$'s, which appear as spatially concentrated ``blobs", as shown in Figure~\ref{fig:events}. To differentiate $\alpha$ and $\beta$ events, we calculate $f_{\textrm{pix}}$, the fraction of a cluster's number of pixels over the total pixel area of the the smallest rectangle drawn around the entire cluster. The value of $f_{\textrm{pix}}$ is larger for $\alpha$'s than high-energy $\beta$'s since $\alpha$'s tend to fill more of the space inside this rectangle. We define $\alpha$'s as clusters having $f_{\textrm{pix}} > 0.75$ for $E > 550$~keV, $f_{\textrm{pix}} > 0.45$ for $E > 900$~keV, or $E > 5$~MeV with linear interpolation between these points, as shown in Figure~\ref{fig:discrimination}~(top). This cut is chosen to accept $>$99.9$\%$ of simulated $\beta$ decays and categorizes all reconstructed events in data with energy greater than 2~MeV as $\alpha$'s. 

Furthermore, $\alpha$'s can be separated into two types by the diffusion of charges: ``plasma" and ``bloomed". Plasma $\alpha$'s correspond to highly-diffuse and round clusters as a result of the plasma effect \cite{Plasma} and originate in the bulk or close to the back of the CCD. Bloomed $\alpha$'s, by contrast, originate near the front of the CCD and produce elongated tracks in $\hat{y}$ due to charge spilling between pixels along a column. More information on both types of $\alpha$'s can be found in Refs.~\cite{Contamination,Plasma}. 
To differentiate plasma and bloomed $\alpha$'s, we determine the spatial RMS spread of a cluster's $x$ and $y$ pixel distributions, $\sigma_{x}$ and $\sigma_{y}$. 
The ratio $\sigma_{x}/\sigma_{y}$ successfully separates $\alpha$'s into the two categories, as shown in Figure~\ref{fig:discrimination} (bottom). Plasma $\alpha$'s are defined to have $\sigma_{x}/\sigma_{y} > 0.9$; bloomed $\alpha$'s are defined to have $\sigma_{x}/\sigma_{y} \leq 0.9$.

Figure~\ref{fig:discrimination} shows two distinct populations of bloomed and plasma $\alpha$'s with energies clustered around 5.3\,MeV with a tail toward lower energies.
These events are consistent with $^{210}$Po decays on the front and back surfaces of the CCDs, which deposit a fraction of their energy in the inactive, few-$\mu$m thick CCD dead layers.
This background arises from long-lived $^{210}$Pb surface contamination due to exposure of the CCDs to environmental $^{222}$Rn during fabrication, packaging, and handling.
We leave a more detailed analysis of surface $^{210}$Pb to a follow-up publication, and instead focus on setting upper limits on bulk $^{210}$Pb contamination. 
Higher energy $\alpha$'s ($E_{\alpha} \gtrsim 5.3$~MeV) must originate from the $^{238}$U and $^{232}$Th chains.
However, the absence of any definitive bulk $^{238}$U and $^{232}$Th coincidences in the data suggests that these events are likely from dust particulates in the CCD surfaces or radon decays in the volume around the CCDs.


\begin{table}[t]
\begin{center}
\caption{Constraints on the activities $A$ of the $^{32}$Si, $^{238}$U, and $^{232}$Th decay chains according to the background-subtracted number of events $n_{\textrm{ev}}$ found in each search with efficiency $\epsilon$ from DAMIC data. Events must meet the correct $\alpha$ or $\beta$ classification for each search (according to Figure~\ref{fig:discrimination}) and pass all relevant energy and time cuts defined here (where $\Delta t$ is defined as the difference between image end times). Note that rapid, subsequent $\alpha$ decays will be reconstructed as a single $\alpha$-like event of higher energy. Additional details on the numbered $\beta \rightarrow \beta$ searches can be found in Table~\ref{searches}. The weighted-average result of the $^{32}$Si searches is given in this table for simplicity, but the individual searches determine $A$ of $120 \pm 30$ $\mu$Bq/kg (1a) and $260 \pm 80$~$\mu$Bq/kg~(1b). Any isotope with $t_{1/2} < 30$ days is assumed to have the activity of its long-lived parent. We truncate the $^{232}$Th chain at $^{212}$Pb as the remaining decays to stable $^{208}$Pb are short-lived ($t_{1/2}~\leq~1$~hr), in secular equilibrium (s.e.) with $^{228}$Th, and not used for this analysis. Decay information is taken from Refs.~\cite{nucleardata,nucleardata2,nucleardata3}.}
\smallskip
\adjustbox{max width=\textwidth}{
\begin{tabular}{ | r c r@{\hskip 0.2in} r@{\hskip 0.1in} r@{\hskip 0.1in} c@{\hskip 0.1in} c@{\hskip 0.1in} c@{\hskip 0.1in} c@{\hskip 0.1in} | c | }
\hline 
\rule{0pt}{2.5ex} & Decay & \multicolumn{1}{c}{$t_{1/2}$} & \multicolumn{1}{c@{\hskip 0.1in}}{Q value} & \multicolumn{1}{c@{\hskip 0.1in}}{Search} & \multicolumn{1}{c@{\hskip 0.1in}}{Energy cut [keV]} & \multicolumn{1}{c@{\hskip 0.1in}}{Time cut [d]} & \multicolumn{1}{c@{\hskip 0.1in}}{$\epsilon$} & \multicolumn{1}{c@{\hskip 0.1in}}{$n_{\textrm{ev}}$} & \multicolumn{1}{c@{\hskip 0.1in} |}{$A$ [$\mu$Bq/kg]} \\ \hline
\multicolumn{1}{| l@{\hskip -0.25in}}{\rule{0pt}{3.0ex} \bf $^{32}\textrm{Si}$ Chain} & \multicolumn{9}{c |}{} \\
\hline
\rule{0pt}{2.5ex} $^{32}\textrm{Si}$ & $\beta$ & 153 yr & 227 keV & \multirow{2}{*}{$\beta \rightarrow \beta \begin{cases} \text{ } \end{cases}$ \hspace{-15pt}}(1a) & $70 < E_{\beta 1} < 230$ & $\Delta t < 70$ & 0.279 & 19.5 & \multirow{2}{*}{$140 \pm 30$}\\  
$^{32}\textrm{P}$ & $\beta$ & 14.3 d & 1.71 MeV & (1b) & $0.5 < E_{\beta 1} < 70$ & 25 $<\Delta t < $ 70 & 0.088 & 12.7 & \\ \hline 
\multicolumn{1}{| l@{\hskip -0.25in}}{\rule{0pt}{3.0ex} \bf $^{238}\textrm{U}$ Chain} & \multicolumn{9}{c |}{}  \\
\hline
\rule{0pt}{2.5ex}$^{238}\textrm{U}$ & $\alpha$ & 4.47 Gyr & 4.27 MeV &  & \multirow{ 2}{*}{$3800 < E_{\alpha 1} < 4600$} & \multirow{ 3}{*}{$\Delta t < 120$} &  \multirow{ 3}{*}{0.650} & \multirow{ 3}{*}{-0.2} &  \multirow{ 3}{*}{$< 11$} \\  
$^{234}\textrm{Th}$ & $\beta$ & 24.1 d & 274 keV & \multicolumn{1}{c@{\hskip 0.1in}}{$\alpha \rightarrow \beta$} & \multirow{ 2}{*}{$E_{\beta 2} > 0.5$} & & & & \\ 
$^{234m}\textrm{Pa}$ & $\beta$ & 1.16 min & 2.27 MeV &  &  &  &  &  & \\ \cline{0-8}
\rule{0pt}{2.5ex}$^{234}\textrm{U}$ & $\alpha$ & 246 kyr & 4.86 MeV & \multicolumn{1}{c@{\hskip 0.1in}}{$^{238}\textrm{U}$ (s.e.)}  &  &  &  &  &  \\ \hline
\rule{0pt}{2.5ex}$^{230}\textrm{Th}$ & $\alpha$ & 75.4 kyr & 4.77 MeV &  &  &  &  &  &  no limit \\ \hline 
\rule{0pt}{2.5ex}$^{226}\textrm{Ra}$ & $\alpha$ & 1.60 kyr & 4.87 MeV &  &  &  &  &  & \\ \cline{0-8}
\rule{0pt}{2.5ex}$^{222}\textrm{Rn}$ & $\alpha$ & 3.82 d & 5.59 MeV &  & \multirow{ 3}{*}{$E_{\alpha} > 15000$} & \multirow{ 3}{*}{$\Delta t =0$} &  \multirow{ 5}{*}{$\sim 1$} & \multirow{ 5}{*}{0} & \multirow{ 5}{*}{$< 5.3$} \\   
$^{218}\textrm{Po}$ & $\alpha$ & 3.10 min & 6.11 MeV & \multicolumn{1}{c@{\hskip 0.1in}}{($\alpha+\alpha+\alpha$)} &  &  &  &  & \\ 
$^{214}\textrm{Pb}$ & $\beta$ & 27.1 min & 1.02 MeV &  \multicolumn{1}{c@{\hskip 0.1in}}{OR}  & \multirow{ 3}{*}{$E_{\alpha 1} > 10000$} & \multirow{ 3}{*}{$\Delta t = t_{\textrm{exp}}$} & & & \\     
$^{214}\textrm{Bi}$ & $\beta$ & 19.9 min & 3.27 MeV & \multicolumn{1}{c@{\hskip 0.1in}}{($\alpha+\alpha) \rightarrow \beta/\alpha$} &  &  &  &  & \\ 
$^{214}\textrm{Po}$ & $\alpha$ & 164 $\mu$s & 7.83 MeV &  &  &  &  &  &\\ \hline  
\rule{0pt}{2.5ex}$^{210}\textrm{Pb}$ & $\beta$ & 22.2 yr & 63.5 keV & \multicolumn{1}{c@{\hskip 0.1in}}{\multirow{ 2}{*}{$\beta \rightarrow \beta$ (2)}} & \multirow{ 2}{*}{$0.5 < E_{\beta 1} < 70$} & \multirow{ 2}{*}{$\Delta t < $ 25} & \multirow{ 2}{*}{0.734} & \multirow{ 2}{*}{47.1} & \multirow{ 2}{*}{$< 160$}\\
$^{210}\textrm{Bi}$ & $\beta$ & 5.01 d & 1.16 MeV &  &  &  &  &  &\\ \cline{0-8}
\rule{0pt}{2.5ex}$^{210}\textrm{Po}$ & $\alpha$ & 138 d & 5.41 MeV &  &  &  &  &  & \\ \hline 
\multicolumn{1}{| l@{\hskip -0.25in}}{\rule{0pt}{3.0ex} \bf $^{232}\textrm{Th}$ Chain} & \multicolumn{9}{c |}{}  \\
\hline
\rule{0pt}{2.5ex}$^{232}\textrm{Th}$ & $\alpha$ & 14.0 Gyr & 4.08 MeV & \multicolumn{1}{c@{\hskip 0.1in}}{$^{228}\textrm{Th}$ (s.e.)} &  &  &  &  & $<7.3$\\ \hline  
\rule{0pt}{2.5ex}$^{228}\textrm{Ra}$ & $\beta$ & 5.75 yr & 45.5 keV & \multicolumn{1}{c@{\hskip 0.1in}}{\multirow{ 2}{*}{$\beta \rightarrow \beta$ (3)}}  & \multirow{ 2}{*}{$0.5 < E_{\beta 1}< 55$} & \multirow{ 2}{*}{$\Delta t < $ 1.3} & \multirow{ 2}{*}{0.440} & \multirow{ 2}{*}{2.6} & \multirow{ 2}{*}{$< 40$}\\
$^{228}\textrm{Ac}$ & $\beta$ & 6.15 hr & 2.12 MeV &  &  &  &  &  &\\ \hline
\rule{0pt}{2.5ex}$^{228}\textrm{Th}$ & $\alpha$ & 1.91 yr & 5.52 MeV & \multicolumn{1}{c@{\hskip 0.1in}}{\multirow{ 4}{*}{$\alpha \rightarrow (\alpha+\alpha)$}} & \multirow{ 4}{*}{$E_{\alpha 2} > 10000$} & \multirow{ 4}{*}{$\Delta t = t_{\textrm{exp}}$} & \multirow{ 4}{*}{0.727} & \multirow{ 4}{*}{0} & \multirow{ 4}{*}{$<7.3$} \\
$^{224}\textrm{Ra}$ & $\alpha$ & 3.63 d & 5.79 MeV &  &  &  &  &  &\\  
$^{220}\textrm{Rn}$ & $\alpha$ & 55.6 s & 6.40 MeV &  &  &  &  &  &\\   
$^{216}\textrm{Po}$ & $\alpha$ & 145 ms & 6.91 MeV &  &  &  &  &  &\\  \cline{0-8}  
\rule{0pt}{2.5ex}$^{212}\textrm{Pb}$ & $\beta$ & 10.6 hr & 569 keV &  &  &  &  &  &
\vspace{-2.5pt}\\
... & & & & & & & & & \\
\hline 
\end{tabular}}
\label{decays}
\end{center}   
\end{table}


\section{Intrinsic $^{32}$Si Measurement}

We leverage the unique spatial resolution of DAMIC CCDs to search for decays coming from the same decay chain at the same location over long times. We conduct searches to identify spatially-correlated decay sequences in order to directly measure bulk radioactive contamination in the CCDs. The isotope $^{32}$Si is produced by spallation, or nuclear fragmentation, of atmospheric $^{40}$Ar by impact of cosmic rays; it falls to the surface with precipitation and is present when silicon is eventually gathered to fabricate ingots for semiconductor production~\cite{Silicon}. The isotope $^{32}$Si and its daughter, $^{32}$P, are of particular concern for DAMIC and other silicon-based direct-detection experiments, because they are intrinsic to the bulk of the detector with $\beta$ electrons that span the low energies of interest for DM searches. The sensitivity of next-generation, large-exposure, solid state silicon experiments~\cite{Sensei,DAMIC-M,SuperCDMS} depends greatly on how much intrinsic radioactive contamination can be reduced. Improving understanding of such contamination aides in establishing mitigation protocols for the material selection and fabrication of future detectors.

The decay characteristics, specifically the half-life ($t_{1/2}$) and Q values of isotope decays, listed in Table~\ref{decays}, guide the searches. For a decay sequence to be considered a candidate: events must occur in the same CCD, isotope daughter clusters have to occur in later images than their parents, clusters must match appropriate $\alpha$, $\beta$ classification and be spatially-coincident (minimum one-pixel overlap) in unmasked regions of the CCD, and sequences have to fall within energy ($E$) and separation time ($\Delta t$) selection, as defined in Table~\ref{decays}. We constrain the separation time of most searches to be within $\approx 5 t_{1/2}$.

\begin{table}[t]
\begin{center}
\caption{Details for the $\beta \rightarrow \beta$ coincidence searches, including pair selection and time efficiencies ($\epsilon_{\textrm{sel}}$ and $\epsilon_{t}$, respectively), number of coincident pairs found ($N_{\textrm{pair}}$), number of accidental pairs expected ($N_{\textrm{acc}}$), and number of pairs expected from overlap with other decay searches ($N_{\textrm{overlap}}$), which are used to calculate the adjusted number events ($n_{\mathrm{ev}}$) attributed to coincident decays for a given search.}
\smallskip
\begin{tabular}{ | r@{\hskip 0.1in} c@{\hskip 0.1in} c@{\hskip 0.1in} c@{\hskip 0.1in} c@{\hskip 0.1in} c c | } \hline 
\rule{0pt}{2.5ex}Isotope & $\epsilon_{\textrm{sel}}$ & $\epsilon_{t}$ & $N_{\textrm{pair}}$ & $N_{\textrm{acc}}$ & $N_{\textrm{overlap}}$ & $n_{\textrm{ev}}$ \\ \hline 
\rule{0pt}{2.5ex}\multirow{2}{*}{$^{32}$Si }(1a) & 0.398 & 0.701 & 26 & $6.5 \pm 0.1$ & 0 & $19.5 \pm 5.1$ \\ 
(1b) & 0.521 & 0.169 & 17 & $3.0 \pm 0.1$ & $1.3 \pm 0.3$ & $12.7 \pm 4.1$ \\ 
\rule{0pt}{3.5ex}$^{210}$Pb (2) & 0.981 & 0.748 & 69 & $2.5 \pm 0.1$ & $19.4 \pm 5.1$ & $27.1 \pm 9.7$ \\ 
\rule{0pt}{3.5ex}$^{228}$Ra (3) & 1 & 0.440 & 8 & $0.08 \pm 0.01$ & $5.28 \pm 1.15$ & $2.6 \pm 3.0$\\ \hline 
\end{tabular}
\label{searches}
\end{center}   
\end{table}

\begin{figure}[t]
\begin{center}
\includegraphics[width=0.7\textwidth, trim=0 0 0 0, clip=true]{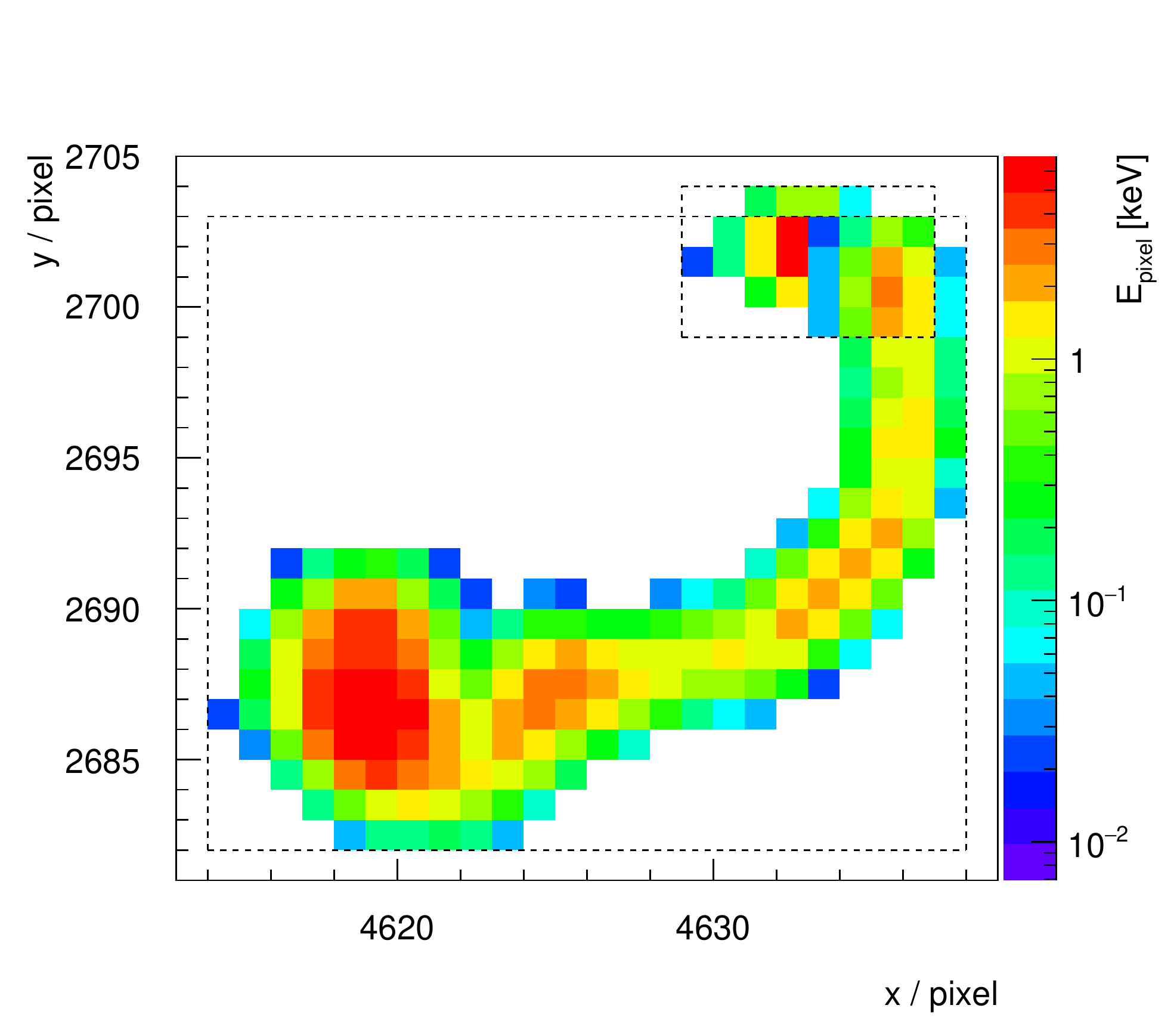}
\end{center}
\caption{A spatially-correlated $\beta \rightarrow \beta$ candidate decay in the reconstructed data from the $^{32}$Si-$^{32}$P search. The first decay has an energy of E$_{\beta_{1}}$ = 77~keV and the second decay has an energy of E$_{\beta_{2}}$ = 399~keV with a time separation of $\Delta$t~=~13.7~d.}
\label{fig:cluster}
\end{figure}

The analysis of $^{32}$Si-$^{32}$P spatially-coincident $\beta \rightarrow \beta$ events is split into two searches according to Table~\ref{decays}, to disentangle any event overlap with spatially-coincident $^{210}$Pb-$^{210}$Bi decays. The first $^{32}$Si search (1a) is at an energy range above the Q value of $^{210}$Pb, and thus includes no potential event overlap. The second $^{32}$Si search (1b) includes a lower separation time bound of 25~d ($\approx 5 t_{1/2}$ of $^{210}$Pb but only $1.75 t_{1/2}$ of $^{32}$Si). Figure~\ref{fig:cluster} shows an example of a spatially-correlated candidate decay from the $^{32}$Si search. We additionally construct a $^{210}$Pb-$^{210}$Bi $\beta \rightarrow \beta$ search (2) to extract an upper limit on the isotopic contamination of the bulk silicon, which is expected to be out of secular equilibrium with its parent $^{238}$U. 
Finally, we construct a $^{228}$Ra-$^{228}$Ac $\beta \rightarrow \beta$ search (3) to constrain the part of the $^{232}$Th chain which may be out of secular equilibrium.
Searches (2) and (3) will be discussed below along with the rest of the $^{238}$U and $^{232}$Th decay chains.

We adjust the number of identified candidate sequences ($N_{\textrm{pair}}$) to account for accidental spatial coincidences of clusters ($N_{\textrm{acc}}$) and overlapping events from searches of other decay chains ($N_{\textrm{overlap}}$). The number of accidental spatial coincidences is simulated by randomizing cluster positions from data over many iterations, and reapplying the search criteria for each iteration. One thousand iterations are performed to obtain the mean value of accidental coincidences, as reported for $\beta \rightarrow \beta$ searches in Table~\ref{searches}. The number of events listed in Table~\ref{decays} reflect the number of identified sequences adjusted for accidentals and overlap ($n_{\textrm{ev}} = N_{\textrm{pair}} - N_{\textrm{acc}} - N_{\textrm{overlap}}$).

Pair selection efficiencies ($\epsilon_{\textrm{sel}}$)\textemdash efficiencies of matching spatially-correlated parent-daughter sequences passing energy cuts\textemdash are calculated using \texttt{GEANT4} ~\cite{geant4}, with 10,000 decays individually simulated for each isotope of interest. Steps of image processing and clustering reconstruction are applied to each simulation output file, to ensure that the same set of clustering parameters as extracted for DAMIC images are present. There were breaks in data acquisition due to image readout ($<9$~min per image) and power outages at the underground site. As such, the time efficiency ($\epsilon_{t}$) of each search is obtained by analytically solving for probabilities of seeing subsequent decays within the time cuts given these periods of downtime, following identification of initial (first-decay) candidates. Time efficiency values are cross-checked with Monte Carlo simulations, which agree with analytical calculations to within $< 1$\%.  The time and pair selection efficiencies of $^{32}$Si and $^{210}$Pb searches can be found in Table~\ref{searches}. The overall efficiency of a search is equal to the product of the individual efficiencies ($\epsilon = \epsilon_{\textrm{sel}} \times \epsilon_{t}$), and is used to calculate the isotopic activity $A$, reported in Table~\ref{decays}, such that
\begin{equation} \label{eq:eq1}
A = \frac{n_{\textrm{ev}}}{\epsilon Mt},
\end{equation}
where $Mt$ is the total exposure considered in the search, 6.5 kg-d for this analysis. 

We measure a bulk $^{32}$Si activity of $140 \pm 30$ $\mu$Bq/kg ($\pm 1 \sigma$ statistical uncertainty) by taking the weighted average of the split searches in Table~\ref{decays} ($A_{\textrm{1a}} = 120~\pm~30$~$\mu$Bq/kg; $A_{\textrm{1b}} = 260~\pm~80$~$\mu$Bq/kg). When calculating the activity for search (1b) according to Eq.~\ref{eq:eq1}, the $1.3 \pm 0.3$ overlapping coincident events in Table~\ref{searches} come from \text{$^{210}$Pb-$^{210}$Bi} coincident decays. The ($\times$6) smaller central value of $^{32}$Si contamination compared to DAMIC's previous R\&D measurement of $926^{+1273}_{-752}$ $\mu$Bq/kg (95\%~CI)~\cite{Contamination} suggests that intrinsic contamination levels may vary depending on the silicon sample~\cite{Silicon}, as the CCDs in the previous study were fabricated from a different silicon ingot.

\section{Uranium and Thorium Chains}

We construct similar searches to look for the activities of $^{238}$U and $^{232}$Th chains in the bulk silicon using both $\alpha$ and $\beta$ decays. 
Parts of the $^{238}$U and $^{232}$Th chains include multiple subsequent decays with half-lives much less than an image exposure time and should result in spatially-coincident events within the same image. 
For such sequences of short-lived bulk isotopes, we can assume a selection efficiency $\epsilon_{\textrm{sel}} \approx 1$ to reconstruct the rapid decays as a single cluster.  In certain cases, we can use the searches for these short-lived isotopes to place limits on parent isotopes or any longer-lived isotopes immediately preceding them. 

\subsection{$^{238}$U}
We begin with the $^{238}$U decay chain, for which we perform a number of searches. We constrain the activity of $^{238}$U by performing an $\alpha \rightarrow \beta$ search. If an $\alpha$ parent candidate corresponding to the $^{238}\textrm{U}$-$^{234}\textrm{Th}$ decay were observed, we would then expect a $\beta$ decay from $^{234}$Th ($t_{1/2}$ = 24 d), and an additional $\beta$ from the rapid decay of $^{234m}$Pa ($t_{1/2}$ = 1.2 min). As before, we calculate the time efficiency for this search analytically and through a Monte Carlo simulation; we obtain a value of $\epsilon_{t}$~=~0.650. Observation of a single $\alpha \rightarrow \beta$ sequence in our data is consistent with $N_{\textrm{acc}} = 1.18 \pm 0.03$ and allows us to set an upper limit (95\% CL) on $^{238}$U contamination of $< 11$ $\mu$Bq/kg ($0.9$ ppt). 

\subsection{$^{222}$Rn}
Further down the $^{238}$U decay chain is $^{222}$Rn ($t_{1/2}$ = 3.8 d), a noble element with high mobility produced from the decay of $^{226}$Ra. 
Following the primary $^{222}$Rn $\alpha$ , we would expect an $\alpha$ from $^{218}$Po within minutes and energy pile-up from further short-lived isotopes (the decays of \text{$^{214}$Pb-Bi-Po}) within hours. As such, we construct a search for pile-up events within a single image with total cluster energy $E >$ 15 MeV, or a first pile-up of events with $E >$ 10 MeV followed by a spatially-coinciding $\alpha$ (or $\alpha+\beta$) in the next image. While two parent candidates with 10~MeV~$<~E~<$~15~MeV are found, there are no spatially-coinciding daughter decays in the following images. No clusters with $E>15$~MeV are observed. Thus, we place a limit of $< 5.3$ $\mu$Bq/kg (95$\%$ CL) for $^{222}$Rn in the bulk of the CCDs. 
The diffusion coefficient of elemental gases in crystalline silicon at cryogenic temperatures is expected to be $\ll 10^{-2}~\mu$m$^2$~d$^{-1}$, based on measurements of xenon at 700$^{\circ}$C~\cite{radon,radon2}. 
Thus, we do not expect $^{222}$Rn to readily escape the CCDs, and the $^{222}$Rn measurement can be considered an upper limit on the $^{226}$Ra activity in the CCDs.

\subsection{$^{210}$Pb}
At the bottom of the $^{238}$U decay chain is $^{210}$Pb ($t_{1/2}$ = 22.2 yr), which is expected to be out of secular equilibrium with the rest of the $^{238}$U decay chain, and whose $\beta$ spectrum is a major concern for DM searches.
We implement a $\beta \rightarrow \beta$ search similar to $^{32}$Si to find coincident $^{210}$Pb-$^{210}$Bi decays. Unlike the case of $^{32}$Si, which is expected to be intrinsic to the CCD bulk, the location of $^{210}$Pb contamination is unknown. The selection efficiency of the $^{210}$Pb search is highly dependent on the location of the $^{210}$Pb contamination, whether it is in the bulk silicon or distributed on the CCD surfaces. In this analysis, we make the overly-conservative assumption that all identified $^{210}$Pb coincidences are in the bulk of the silicon in order to place a strong upper limit on such contributions of $< 160$ $\mu$Bq/kg (95$\%$ CL). 
This upper limit is orders of magnitude better than the mBq/kg sensitivity obtained by direct assay techniques that measure bulk $^{210}$Pb in materials~\cite{icpms,Assay2}. 
When calculating this activity according to Eq.~\ref{eq:eq1}, the $19.4 \pm 5.1$ overlapping coincident events in Table~\ref{searches} come from $^{32}$Si-$^{32}$P coincident decays with intersecting energy ranges, calculated using the activity from the $^{32}$Si search (1a). 
We expect the majority of the 69 coincident $\beta \rightarrow \beta$ candidates in this search to be from surface $^{210}$Pb, as suggested by the two distinct $\alpha$ populations observed in Figure~\ref{fig:discrimination}.

\subsection{$^{232}$Th/$^{228}$Th}
Because $^{232}$Th and $^{228}$Th are two long-lived isotopes of the same element, we assume that they remain in secular equilibrium throughout the production of high-purity detector-grade silicon. We perform an $\alpha \rightarrow \alpha$ spatial coincidence search across different images. We search first for the $\alpha$ decay of $^{228}$Th followed by (in a separate image) the rapid sequence of $\alpha$ decays from $^{224}$Ra ($t_{1/2}$ = 3.7 d), $^{220}$Rn ($t_{1/2}$ = 56 s), and $^{216}$Po~($t_{1/2}$~=~145 ms). We assume that thoron ($^{220}$Rn) does not diffuse significantly in its minute-long lifetime. For the second cluster of this search, we thus expect a triple-$\alpha$ pileup with an energy exceeding 18 MeV within a single image. We construct an $\alpha \rightarrow \alpha$ search in which the energy of the second cluster $E > 10$ MeV. 
The time efficiency of the search, dominated by the 3.7~d half-life of $^{224}$Ra, is $\epsilon_{t} = 0.727$. Non-observation of appropriate $\alpha$ sequences allows us to place an upper limit (95\% CL) on $^{228}$Th, and by extension $^{232}$Th, contamination: $< 7.3$~$\mu$Bq/kg (1.8 ppt).              

\subsection{$^{228}$Ra}
Another long-lived decay in the $^{232}$Th chain is $^{228}$Ra, which can be identified with a $\beta~\rightarrow~\beta$ search. We place an energy cut on the initial $\beta$ of $E_{\beta 1} < 55$ keV. Given the 0.254~d half-life of $^{228}$Ac, we obtain $\epsilon_{t}$ = 0.440 that the two $\beta$'s occur in different images. Eight candidate events are identified, all of which also appear in the $\beta~\rightarrow~\beta$ search of $^{210}$Pb. 
Applying the time and energy bounds of the $^{228}$Ra search to bulk $^{210}$Pb decay gives overlap efficiencies of $\epsilon_{\textrm{sel}} = 0.802$ and $\epsilon_{t} = 0.11$, resulting in an expectation of 5.28 $\pm$ 1.15 overlapping events. 
We note that spatial accidentals are very small ($0.08$~events) for this search given the small value of $t_{1/2}$. 
We use the observation of $n_{\textrm{ev}} = 2.6 \pm 3.1$ events to place an upper limit on $^{228}$Ra: $<40$ $\mu$Bq/kg (95$\%$ CL). 

\section{Conclusions and Outlook}
We demonstrate an analysis technique that utilizes spatial-coincidence searches to directly measure the radioactive contamination within DAMIC CCDs. We determine the activity of $^{32}$Si in our CCDs, and set limits on bulk contamination from the $^{238}$U and $^{232}$Th chains, including the activities of all $\beta$ emitters, problematic backgrounds for DM searches. 
The $^{32}$Si result, together with ICP-MS, germanium $\gamma$-counting, \texttt{GEANT4} simulations, and detector component activation analysis, plays a critical role in constraining the background model for DAMIC's WIMP search analysis~\cite{damic2020}. 
It provides the first comparison of the $^{32}$Si contamination level in different science-grade silicon detectors, and suggests that $^{32}$Si levels may vary geographically, as suggested in Ref.~\cite{Silicon}.
CCDs will be critical in the effort to scale up silicon-based detectors as a feasible technology to screen detector-grade silicon for the desired background levels of next-generation experiments. 
This may allow for low background silicon ingot selection via multiple production cycles across several ingots. 
Both the future CCD program~\cite{DAMIC-M,DAMIC-M2} and other leading silicon-based experiments~\cite{SuperCDMS} benefit from such measurements of intrinsic contamination. 
\\

\acknowledgments We thank Richard N. Saldanha for useful discussion regarding the diffusion of radon in crystalline
silicon. 

We are grateful to SNOLAB and its staff for support through underground space, logistical and technical services. SNOLAB operations are supported by the Canada Foundation for Innovation and the Province of Ontario Ministry of Research and Innovation, with underground access provided by Vale at the Creighton mine site. The CCD development work was supported in part by the Director, Office of Science, of the U.S. Department of Energy under Contract No. DE-AC02-05CH11231. We acknowledge financial support from the following agencies and organizations: National Science Foundation through Grant No. NSF PHY-1806974 and Kavli Institute for Cosmological Physics at The University of Chicago through an endowment from the Kavli Foundation; 
Fermi National Accelerator Laboratory (Contract No. DE-AC02-07CH11359); Institut Lagrange de Paris Laboratoire d’Excellence (under Reference No. ANR10-LABX-63) supported by French state funds managed by the Agence Nationale de la Recherche within the Investissements d'Avenir program under Reference No. ANR-11-IDEX-0004-02; Swiss National Science Foundation through Grant No. 200021 153654 and via the Swiss Canton of Zurich; Mexico’s Consejo Nacional de Ciencia y Tecnolog\'ia (Grant No. 240666) and Direcc\'ion General de Asuntos del Personal Acad\'emico–Universidad Nacional Aut\'onoma de M\'exico (Programa de Apoyo a Proyectos de Investigaci\'on e Innovaci\'on Tecnol\'ogica Grants No. IB100413 and No. IN112213); STFC Global Challenges Research Fund (Foundation Awards Grant ST/R002908/1).

\end{document}